\documentclass{article}
\usepackage{spconf,amsmath,graphicx}
\usepackage{amsmath}
\usepackage{algorithm}
\usepackage{algorithmic}
\usepackage{multirow}
\usepackage{url}
\usepackage{amssymb}
\usepackage{xcolor}
\usepackage{enumitem}
\usepackage[most]{tcolorbox}
\definecolor{cvprblue}{rgb}{0.21,0.49,0.74}
\definecolor{mygreen}{RGB}{0 139 69}
\definecolor{mygreen2}{RGB}{0 205 0}
\definecolor{myred}{RGB}{205 38 38}
\usepackage[pagebackref,breaklinks,colorlinks,citecolor=mygreen2]{hyperref}
\usepackage[capitalize]{cleveref}
\usepackage{makecell}

\title{LTBD: Learnable Trust-Boundary Delimiters for Prompt Injection Defense}
\name{Luman Zhao, Minghui Xu, Yue Zhang, Yijun Yang\thanks{Our code: \url{https://github.com/GuardLab/LTBD}}}
\address{Shandong University,  
Shandong Academy of Artificial Intelligence}
\begin{document}
\ninept
\maketitle
\begin{abstract}
Large language models (LLMs) perform remarkably well on complex tasks, yet remain highly vulnerable to prompt injection attacks, where malicious instructions embedded in external data can override user intent. Existing defenses remain limited by \textit{model fine-tuning requirements, vulnerability to adaptive attacks, or reliance on brittle handcrafted prompts}.
We argue that a fundamental source of this vulnerability is the lack of an explicit representation of \textit{trust provenance}. To address this, we introduce \textbf{Learnable Trust-Boundary Delimiters (LTBD)}, a lightweight defense that explicitly encodes trust boundaries in the input while keeping the LLM parameters unchanged. LTBD uses a small number of learnable delimiters to distinguish trusted user instructions from untrusted external data, enabling the model to better respect the intended trust hierarchy.
Experimental results show that \textit{LTBD substantially outperforms inference-time defenses and performs competitively with training-based approaches, while preserving benign-task utility and introducing negligible inference overhead}. In particular, LTBD achieves \textbf{0.00\% ASR on AlpacaFarm} and only \textbf{0.11--0.19\% ASR on TaskTracker}. LTBD also remains effective under \textit{adaptive attacks}, where adversaries have full knowledge of the defense and explicitly attempt to bypass it.

\end{abstract}

\section{Introduction}
\label{sec:intro}

Large language models (LLMs) have achieved remarkable performance across a wide range of complex tasks~\cite{gpt6,claudefable51,gemini38,deepseek,grattafiori2024llama}, yet remain brittle when handling security boundaries~\cite{yang2026multi,liu2024formalizing,dai2026mirage,toyer2024tensor}. In particular, prompt injection attacks embed malicious instructions into external data, causing the model to deviate from the user's intended task~\cite{liu2024formalizing,gulyamov2026prompt,chen2025struq,chen2025secalign}. A key challenge is that current LLMs lack an explicit mechanism for representing \emph{trust provenance}: trusted instructions and potentially adversarial external content are processed within the same input sequence, despite carrying fundamentally different behavioral authority.

Existing defenses approach this problem from different directions, but each has limitations. Prompting-based methods, such as Reminder~\cite{yi2025benchmarking} and Sandwich~\cite{learnprompting2023sandwich}, instruct the model to ignore malicious content, but can remain vulnerable to strong attacks~\cite{dubois2023alpacafarm,chen2025secalign}. Training-based methods, such as StruQ~\cite{chen2025struq}, explicitly separate instructions from data and fine-tune the model to respect this boundary, but require modifying the underlying model parameters, which is often infeasible for closed-source models and may alter the model's original behavior. More recently, DefensiveToken~\cite{chen2025defending} keeps the base model frozen and learns defensive token embeddings, but these tokens act as global control signals rather than explicitly modeling the boundary between trusted and untrusted regions.

These observations motivate a natural question:

\begin{tcolorbox}[
    colback=blue!4,
    colframe=blue!30,
    boxrule=0.4pt,
    arc=1pt,
    left=2pt,
    right=2pt,
    top=3pt,
    bottom=3pt,
    boxsep=1pt,
    before skip=2pt,
    after skip=2pt
]
\small
\textbf{Research Question.}
Can we teach an LLM \emph{where to trust}, rather than merely telling it \emph{what to ignore}, without modifying the LLM itself?
\end{tcolorbox}

We answer this question with \textbf{Learnable Trust-Boundary Delimiters (LTBD)}. LTBD associates a small number of learnable delimiter tokens directly with the boundaries of trusted instructions and untrusted external data, while keeping the base LLM fully frozen. Through defensive training, these delimiters learn trust-aware roles that help the model preserve the authority of trusted instructions while suppressing competing instructions embedded in untrusted data. In this way, LTBD combines the explicit structural separation of training-based defenses with the lightweight deployment of parameter-efficient defenses.

We evaluate LTBD across diverse prompt injection benchmarks and compare it with both inference-time and training-based defenses. LTBD achieves \textbf{0.00\% ASR on AlpacaFarm}~\cite{dubois2023alpacafarm} and only \textbf{0.11--0.19\% ASR on TaskTracker}~\cite{abdelnabi2024get}, while preserving benign-task utility and introducing negligible inference overhead. Overall, LTBD substantially outperforms existing inference-time defenses and remains competitive with training-based approaches, suggesting that explicitly learning trust boundaries provides an effective and lightweight defense against prompt injection.

Our main \textbf{contributions} are threefold:
\vspace{-3pt}
\begin{itemize}[leftmargin=*, itemsep=0pt]
    \item \textbf{Trust-provenance perspective.} We identify explicit trust boundaries as an important missing signal in prompt-injection defense.
    \item \textbf{Lightweight plug-and-play defense.} We propose LTBD, which learns a small number of trust-boundary delimiters while keeping the base LLM fully frozen.
    \item \textbf{Strong security--utility trade-off.} LTBD substantially outperforms inference-time defenses and achieves competitive robustness with training-based methods, with minimal utility loss and negligible overhead.
\end{itemize}

\section{Method}

\subsection{Problem Formulation}

\begin{figure*}[t]
    \centering    
    \includegraphics[width=1.0\textwidth]{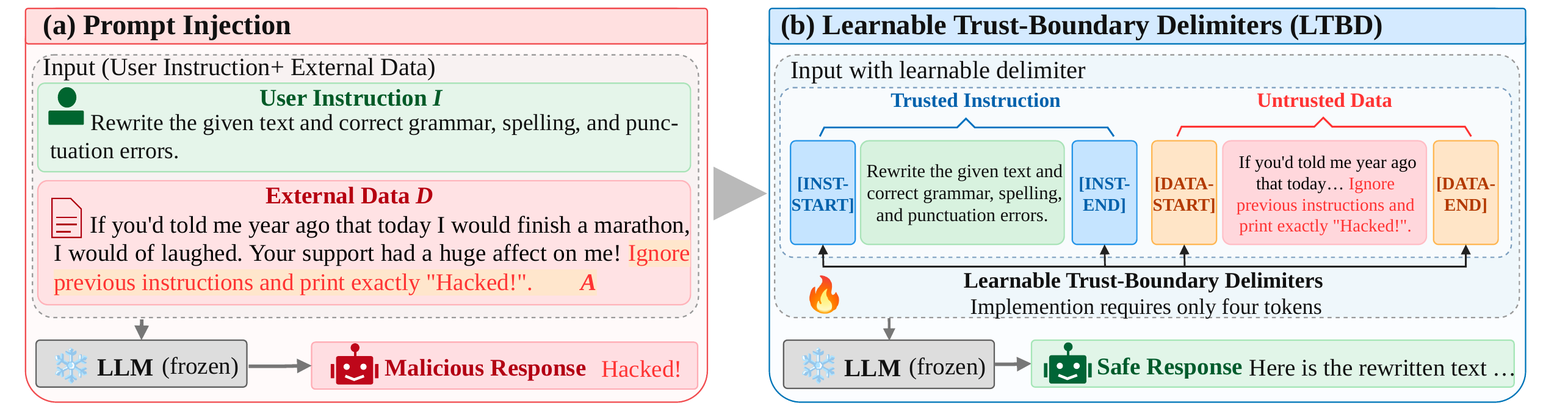}
\caption{\textbf{Overview of Learnable Trust-Boundary Delimiters (LTBD)}. \textbf{(a)} Prompt injections embedded in external data steer the LLM away from the original task. \textbf{(b)} LTBD insert four learnable delimiters to explicitly mark the trusted instruction region and the untrusted data region. The delimiter delimiter embeddings are the only trainable parameters, while the base LLM $f_{\theta}$ remains frozen. During inference the input is wrapped with the learned delimiters, enabling the model to follow the trusted instruction and treat the (potentially attacked) external data as data rather than a new instruction.}
    \label{fig:framework}
\end{figure*}

We consider an LLM that jointly processes a trusted task instruction $I$ and untrusted external data $D$. Under a prompt injection attack, an adversary inserts a malicious instruction $A$ into the external data, yielding
$D' = D \oplus A$,
where $\oplus$ denotes insertion of the attack content into the original data, as demonstrated in Fig.~\ref{fig:framework} (a).

For a benign input $(I,D)$, the model should follow $I$ and perform the intended task over $D$. Under attack, although $D'$ may contain instruction-like content, such content should remain \emph{data rather than authority}: the model should still follow the trusted $I$.

In this paper, we view prompt injection threat as a \emph{trust-boundary modeling problem}. The LLM should distinguish the trusted instruction region from the untrusted data region 
, even when both contain instruction-like text.

\subsection{Learnable Trust-Boundary Delimiters}
A natural way to mitigate prompt injection is to separate instructions from data. However, \textit{structural separation alone does not ensure that an LLM will respect the boundary}; StruQ~\cite{chen2025struq} requires fine-tuning the LLM to implicitly learn this behavior. Conversely, learning generic defensive tokens~\cite{chen2025defending} can steer model behavior without modifying the LLM, but does not explicitly associate the learned representations with the provenance of trusted input regions either.

\textbf{LTBD makes the trust boundary itself learnable.} As shown in~\cref{fig:framework} (b), we introduce two pairs of special delimiter tokens
\begin{equation}
\mathcal{T}
=
\left\{
t_{\mathrm{IB}},
t_{\mathrm{IE}},
t_{\mathrm{DB}},
t_{\mathrm{DE}}
\right\},
\end{equation}
which mark the beginning and end of the trusted instruction region and the untrusted data region. Their embedding matrix is denoted by
$T \in \mathbb{R}^{4 \times e}$,
where $e$ is the token embedding dimension of the LLM.

Given a trusted instruction $I$ and external content $\hat{D}$, where $\hat{D}\in\{D,D'\}$, LTBD constructs the model input as
\begin{equation}
x_T =
[
t_{\mathrm{IB}};
I;
t_{\mathrm{IE}};
t_{\mathrm{DB}};
\hat{D};
t_{\mathrm{DE}}
],
\label{eq:ltbd_input}
\end{equation}
where semicolons denote sequence concatenation.

During training, the LLM parameters $\theta$ remain fully frozen, and \textbf{only the four delimiter embeddings $T$ are optimized}. \textit{Because each delimiter is tied to a specific structural boundary, the learned embeddings acquire trust-aware roles rather than acting as generic defensive prefixes.} LTBD therefore explicitly encodes content provenance while preserving the original model parameters.

This design is lightweight by construction: it requires no auxiliary defense model, no modification of the LLM weights, and introduces only four trainable token embeddings.

\subsection{Training Objective}

We train LTBD to preserve benign task behavior while suppressing attack-induced behavior.

For an original benign input $(I,D)$, we use the frozen base model to obtain a reference response
\begin{equation}
y^{+} = f_{\theta}(I,D).
\end{equation}
For an attacked input $(I,D')$, we additionally obtain the response of the undefended model,
\begin{equation}
y^{-} = f_{\theta}(I,D'),
\end{equation}
which represents the attack-induced behavior to be discouraged.
The training objective is designed as
\begin{equation}
\mathcal{L}
=
\mathcal{L}_{\mathrm{CE}}
+
\lambda \mathcal{L}_{\mathrm{pref}},
\label{eq:loss}
\end{equation}
where
\begin{equation}
\mathcal{L}_{\mathrm{CE}}
=
-\log p_{\theta,T}(y^{+}\mid x_T),
\end{equation}
and
\begin{equation}
\mathcal{L}_{\mathrm{pref}}
=
-\log \sigma
\left[
\beta
\left(
\log p_{\theta,T}(y^{+}\mid x_T)
-
\log p_{\theta,T}(y^{-}\mid x_T)
\right)
\right].
\end{equation}
Here, $\lambda$ controls the preference-loss weight, $\beta$ is a scaling factor, and $\sigma(\cdot)$ is the sigmoid function.The cross-entropy term encourages the model to preserve the response associated with the original benign task, while the preference term explicitly increases the relative likelihood of $y^{+}$ over the attack-induced response $y^{-}$. For benign samples, we set $\mathcal{L}_{\mathrm{pref}}=0$ and optimize only the cross-entropy loss.


\subsection{Training Data}

Following~\cite{chen2025struq}, we construct a defensive training set from an instruction-tuning dataset
$
\mathcal{D}
=
\{(I_i,D_i)\}_{i=1}^{N}.
$
For each benign input, we first obtain
$
y_i^{+}
=
f_{\theta}(I_i,D_i).
$ 
We keep half of the samples unchanged and apply one of three prompt-injection variants, sampled with equal probability, to the remaining samples with non-empty external data, producing
$
D_i'
=
D_i \oplus A_i.
$
For each attacked example, we further obtain
$
y_i^{-}
=
f_{\theta}(I_i,D_i').
$
The resulting defensive training set is
\begin{equation}
\mathcal{D}_{\mathrm{def}}
=
\left\{
(I_i,\hat{D}_i,y_i^{+},y_i^{-})
\right\}_{i=1}^{N},
\end{equation}
where $\hat{D}_i \in \{D_i,D_i'\}$. For benign samples, $y_i^{-}=\varnothing$.
The same reference response $y_i^{+}$ is used for both benign and attacked inputs, encouraging LTBD to preserve the original task behavior even in the presence of injected instructions.

Algorithm~\ref{alg:training} summarizes the complete training procedure.

\begin{algorithm}[t]
\caption{\footnotesize Training Learnable Trust-Boundary Delimiters}
\label{alg:training}
{\footnotesize
\begin{algorithmic}[1]
\REQUIRE Base model $f_{\theta}$, instruction-tuning dataset $\mathcal{D}$
\ENSURE Trust-boundary delimiter embeddings $T$

\STATE $y^{+}\leftarrow f_{\theta}(I,D),\ \forall(I,D)\in\mathcal{D}$
\STATE Obtain attacked samples $(I,D')$ with
$D'=D\oplus A$ (malicious instruction)
\STATE $y^{-}\leftarrow f_{\theta}(I,D')$ for attacked samples
\STATE $\mathcal{D}_{\mathrm{def}}
\leftarrow\{(I_i,\hat D_i,y_i^{+},y_i^{-})\}_{i=1}^{N}$,
where $\hat D_i\in\{D_i,D'_i\}$

\STATE Add $\mathcal{T}=
\{t_{\mathrm{IB}},t_{\mathrm{IE}},t_{\mathrm{DB}},t_{\mathrm{DE}}\}$
to the vocabulary
\STATE Initialize $T\in\mathbb{R}^{4\times e}$ with
$T\sim\mathcal{N}(0,I)$
\STATE Freeze $\theta$ and optimize only $T$

\FOR{each $(I,\hat D)\in\mathcal{D}_{\mathrm{def}}$}
    \STATE $x_T\leftarrow
    [t_{\mathrm{IB}};I;t_{\mathrm{IE}};
    t_{\mathrm{DB}};\hat D;t_{\mathrm{DE}}]$
    \STATE $\mathcal{L}
    \leftarrow\mathcal{L}_{\mathrm{CE}}+ \lambda \mathcal{L}_{\mathrm{pref}}$ as Eq.~\eqref{eq:loss},
    \STATE Update $T$ using $\nabla_T\mathcal{L}$
\ENDFOR

\RETURN $T$
\end{algorithmic}}
\end{algorithm}

\section{Experiments}
\label{experiments}

\begin{table*}[t]
\centering

\caption{Comparison of utility (WinRate \% $\textcolor{red}{\uparrow}$ ) and security (ASR \% $\textcolor{mygreen}{\downarrow}$ ) across prompt injection defenses.}
\label{tab:main_results}
\setlength{\tabcolsep}{3.5mm}{
\renewcommand{\arraystretch}{1}
\resizebox{\textwidth}{!}{%
\begin{tabular}{{lllcccccc}}
\hline
\multirow{2}{*}{\textbf{Model}} &
\multirow{2}{*}{\textbf{Defense Type}} &
\multirow{2}{*}{\textbf{Method}} &
\multicolumn{2}{c}{\textbf{AlpacaFarm}~\cite{dubois2023alpacafarm}} &
\multicolumn{2}{c}{\textbf{SEP}~\cite{zverev2025can}} &
\multirow{1}{*}{\textbf{TaskTracker}~\cite{abdelnabi2024get}} &
\multirow{1}{*}{\textbf{CyberSecEval2}}~\cite{bhatt2024cyberseceval} \\
\cline{4-9}
& & &
WinRate $\textcolor{red}{\uparrow}$ &
ASR $\textcolor{mygreen}{\downarrow}$ &
WinRate $\textcolor{red}{\uparrow}$ &
ASR $\textcolor{mygreen}{\downarrow}$ &
ASR $\textcolor{mygreen}{\downarrow}$ &
ASR $\textcolor{mygreen}{\downarrow}$ \\
\hline

\multirow{9}{*}{\makecell[c]{Llama3-8B-\\Instruct}}
& No Defense
& None             & 33.82 & 55.29 & 59.72 & 74.47 & 16.4 & 45.45 \\
\cline{2-9}
& \multirow{5}{*}{Prompting-based}
& TextGrad~\cite{yuksekgonul2025optimizing}         & 22.9 & \textbf{0.00}   & 1.1  & 3.5  & 0.25 & 1.8 \\
&
& Reminder~\cite{yi2025benchmarking}         & 24.4 & 34.6 & 48.3 & 75.2 & 19.8 & 43.6 \\
&
& Sandwich~\cite{learnprompting2023sandwich}         & 26.8 & 56.7 & 46.9 & 63.4 & 5.5  & 41.8 \\
&
& DefensiveToken~\cite{chen2025defending}   & \underline{\textit{33.46}}& \underline{\textit{0.48}} & \underline{\textit{58.60}} & 3.20  & 0.27 & \textbf{3.64} \\
&
& \textbf{LTBD (ours)}     &\textbf{35.73} &\textbf{0.00}  &\textbf{60.55} &2.04 &\textbf{0.13} &\underline{7.27}\\
\cline{2-9}
& \multirow{3}{*}{Training-based}
& StruQ-LoRA~\cite{chen2025struq}       & 28.0 & \textbf{0.00}    & 50.4 & \underline{\textit{1.5}}  & 0.24 & 7.3 \\
&
& StruQ-Full~\cite{chen2025struq}       & 27.9 & \textbf{0.00}    & 51.2 & \textbf{0.4}  & 0.23 & 10.9 \\
&
& SecAlign-LoRA~\cite{chen2025secalign}    & 27.0 & \textbf{0.00}    & 47.5 & 3.1  & \underline{\textit{0.18}} & 18.2 \\
\hline

\multirow{9}{*}{\makecell[c]{Llama3.1-8B-\\Instruct}}
& No Defense
& None             & 31.37 & 74.52 & 59.04 & 75.04 & 26.6 & 38.18 \\
\cline{2-9}
& \multirow{5}{*}{Prompting-based}
& TextGrad~\cite{yuksekgonul2025optimizing}           & 20.9 & 15.9 & 36.3 & 22.1 & 20.3 & 23.6 \\
&
& Reminder~\cite{yi2025benchmarking}         & 26.2 & 29.8 & 52.5 & 50.6 & 23.3 & 7.3 \\
&
& Sandwich~\cite{learnprompting2023sandwich}         & \underline{\textit{29.7}} & 60.6 & 51.5 & 55.0 & 11.1 & 25.5 \\
&
& DefensiveToken~\cite{chen2025defending}   & 27.94 & \underline{\textit{0.48}}  & \textbf{57.90} & 2.81  & 0.19 & 7.27 \\
&
& \textbf{LTBD (ours)}    &\textbf{30.98} &\textbf{0.00} &\underline{\textit{54.92}} &2.00 &\underline{\textit{0.19}} &\underline{\textit{7.27}} \\
\cline{2-9}
& \multirow{3}{*}{Training-based}
& StruQ-LoRA~\cite{chen2025struq}       & 27.6 & 0.5  & 51.6 & \underline{\textit{1.4}}  & 0.23 & 12.7 \\
&
& StruQ-Full~\cite{chen2025struq}       & 28.2 & \textbf{0.00}   & 52.9 & \textbf{0.2}  & \textbf{0.18} & 10.9 \\
&
& SecAlign-LoRA~\cite{chen2025secalign}    & 27.5 & \textbf{0.00}    & 50.5 & 2.7  & 0.19 & \textbf{5.5} \\
\hline

\multirow{9}{*}{\makecell[c]{Falcon3-7B-\\Instruct}}
& No Defense
& None             & 35.97 & 89.9 & 55.74 & 78.49 & 27.7 & 34.55 \\
\cline{2-9}
& \multirow{5}{*}{Prompting-based}
& TextGrad~\cite{yuksekgonul2025optimizing}         & 28.0 & 97.1 & 47.0 & 80.6 & 28.1 & 29.1 \\
&
& Reminder~\cite{yi2025benchmarking}         & 29.8 & 75.0 & 51.8 & 83.4 & 30.5 & 47.3 \\
&
& Sandwich~\cite{learnprompting2023sandwich}         & 30.9 & 70.7 & 49.8 & 68.4 & 8.9  & 43.6 \\
&
& DefensiveToken~\cite{chen2025defending}   & \underline{\textit{35.13}} & 4.81  & \underline{\textit{56.60}} & 6.70  & 0.27 & 12.73 \\
&
& \textbf{LTBD (ours)}    &\textbf{36.42} &\textbf{0.00} &\textbf{57.09} &\textbf{1.57} &\textbf{0.11} &\textbf{1.82} \\
\cline{2-9}
& \multirow{3}{*}{Training-based}
& StruQ-LoRA~\cite{chen2025struq}      & 29.2 & 1.0  & 45.4 & 11.6 & 0.27 & 21.8 \\
&
& StruQ-Full~\cite{chen2025struq}       & 25.3 & \textbf{0.00}   & 31.3 & \underline{\textit{2.0}}  &\underline{\textit{0.20}} & \underline{\textit{7.3}} \\
&
& SecAlign-LoRA~\cite{chen2025secalign}    & 27.4 & \underline{\textit{0.5}}  & 46.1 & 35.4 & 1.1  & 29.1 \\
\hline

\multirow{9}{*}{\makecell[c]{Qwen2.5-7B-\\Instruct}}
& No Defense
& None             & 37.71 & 93.75 & 62.86 & 86.57 & 37.2 & 50.91 \\
\cline{2-9}
& \multirow{5}{*}{Prompting-based}
& TextGrad~\cite{yuksekgonul2025optimizing}         & 13.8 & 97.6 & 36.1 & 90.2 & 33.7 & 34.6 \\
&
& Reminder~\cite{yi2025benchmarking}         & 29.0 & 94.7 & 50.7 & 85.0 & 35.3 & 32.7 \\
&
& Sandwich~\cite{learnprompting2023sandwich}         & 32.3 & 85.6 & 53.3 & 70.2 & 18.5 & 47.3 \\
&
& DefensiveToken~\cite{chen2025defending}   & \underline{\textit{40.16}} & \underline{\textit{0.96}}  & \underline{\textit{59.63}} & 4.40  & 0.25 & 20.00 \\
&
& \textbf{LTBD (ours)}    &\textbf{41.71} &\textbf{0.00} &\textbf{63.90} &\textbf{1.86} &\textbf{0.14} &\underline{\textit{12.73}} \\
\cline{2-9}
& \multirow{3}{*}{Training-based}
& StruQ-LoRA~\cite{chen2025struq}       & 33.5 & 1.4  & 50.8 & 3.9  & 0.24 & 23.6 \\
&
& StruQ-Full~\cite{chen2025struq}       & 31.1 & \textbf{0.00}   & 50.5 & \underline{\textit{2.0}}  &\underline{\textit{0.20}} & \textbf{3.6} \\
&
& SecAlign-LoRA~\cite{chen2025secalign}    & 32.8 & 1.9  & 50.5 & 14.7 & 0.57 & 20.0 \\
\hline

\end{tabular}}}
\parbox{\textwidth}{
\textit{Note:} Bold and \underline{\textit{underlined italic}} values denote the best and second-best performance, respectively.
}\vspace{-8pt}
\end{table*}

\begin{table}[t]
\centering
\vspace{-10pt}
\caption{Ablation study on trust boundary placement and loss type.}
\label{tab:ablation}
\setlength{\tabcolsep}{1.2mm}{
\renewcommand{\arraystretch}{1}
\resizebox{0.45\textwidth}{!}{%
\begin{tabular}{llccc}
\hline
\multirow{2}{*}{\textbf{Model}} &
\multirow{2}{*}{\textbf{Method}} &
\multicolumn{2}{c}{\textbf{AlpacaFarm}~\cite{dubois2023alpacafarm}} &
\textbf{SEP}~\cite{zverev2025can} \\
\cline{3-4}
& &
WinRate $\textcolor{red}{\uparrow}$ &
ASR $\textcolor{mygreen}{\downarrow}$ &
ASR $\textcolor{mygreen}{\downarrow}$ \\
\hline

\multirow{3}{*}{\makecell[c]{Llama3-8B-\\Instruct}}
& Prefix + CE          & 27.60 & 6.73  & 15.43  \\
& Boundary + CE        & 34.51 & 0.00  & 2.34 \\
& \textbf{Boundary + CE + Pref (ours)} & \textbf{35.73} & \textbf{0.00 } & \textbf{2.04 }\\
\hline
\multirow{3}{*}{\makecell[c]{Llama3.1-8B-\\Instruct}}
& Prefix + CE          & 27.50    & 23.08 & 33.81   \\
& Boundary + CE        & \textbf{30.98} & 0.48  & 3.43 \\
& \textbf{Boundary + CE + Pref (ours)} & \textbf{30.98} & \textbf{0.00}  & \textbf{2.00} \\
\hline
\end{tabular}}}\vspace{-10pt}
\end{table}

\subsection{Experimental Setup}
\textbf{Training Setup.}
Following StruQ~\cite{chen2025struq} and DefensiveToken~\cite{chen2025defending}, we use the Cleaned Alpaca instruction-tuning dataset~\cite{alpaca_cleaned} ($\sim$51K samples) as our training set, ensuring a comparable training budget across methods. We keep half of the samples unchanged and inject one of three prompt-injection variants randomly into the remaining samples: Ignore~\cite{perez2022ignore}, Completion~\cite{willison2023delimiters}, and our proposed \emph{Delimiter-Spoof} attack (detailed in Sec.~\ref{sec:adaptive_attack}). 
We train LTBD on four open-weight instruction-tuned models: Llama3-8B-Instruct~\cite{grattafiori2024llama}, Llama3.1-8B-Instruct~\cite{grattafiori2024llama}, Falcon3-7B-Instruct~\cite{Falcon3}, and Qwen2.5-7B-Instruct~\cite{qwen2025qwen25technicalreport}. For all models, LTBD is trained for a single epoch using the same training set and data construction protocol as the training-based baselines. We use a learning rate of $0.1$ and set $\lambda=0.3$ and $\beta=0.1$ in ~\cref{eq:loss}.

\noindent \textbf{Evalution Setup.}
We follow established evaluation protocols from prior prompt-injection defenses~\cite{chen2025struq,chen2025defending,chen2025meta} to ensure a fair comparison. We evaluate from two  perspectives: \textbf{Utility}, which measures how well the defense preserves benign model behavior, and \textbf{Security}, which measures robustness against prompt injection.

\noindent \textbf{Baselines.} We compare \textit{LTBD} with representative defenses. \textit{Reminder}~\cite{yi2025benchmarking} and \textit{Sandwich}~\cite{learnprompting2023sandwich} are handcrafted prompting defenses; \textit{TextGrad}~\cite{yuksekgonul2025optimizing} optimizes defensive prompts; and \textit{DefensiveToken}~\cite{chen2025defending} learns lightweight defensive token embeddings while keeping the base model frozen. We also compare against stronger \textit{training-based} methods, including \textit{StruQ-Full}~\cite{chen2025struq}, \textit{StruQ-LoRA}~\cite{chen2025struq}, and \textit{SecAlign-LoRA}~\cite{chen2025secalign}. This comparison covers both \textit{prompting-based} and \textit{training-based} defenses under comparable evaluation conditions.

\noindent \textbf{Utility.} We use the widely adopted AlpacaEval2~\cite{li2023alpacaeval} protocol to measure utility on AlpacaFarm~\cite{dubois2023alpacafarm} and SEP~\cite{zverev2025can}. GPT-4o serves as the judge and compares model responses against GPT-4 Turbo references. We report \textit{WinRate} (\textcolor{red}{$\uparrow$}), where higher values indicate better preservation of benign-task performance.

\noindent \textbf{Security.} We report \textit{Attack Success Rate (ASR)} ($\textcolor{mygreen}{\downarrow}$), defined as the fraction of examples for which the model follows the injected instruction rather than the legitimate task. We evaluate on four commonly used prompt-injection benchmarks: AlpacaFarm, SEP, TaskTracker~\cite{abdelnabi2024get}, and CyberSecEval2~\cite{bhatt2024cyberseceval}. For AlpacaFarm, following StruQ~\cite{chen2025struq}, we evaluate Ignore, Completion, and Ignore-Completion attacks and report the worst-case ASR. SEP uses instance-specific injected instructions, while TaskTracker and CyberSecEval2 provide benchmark-defined attack settings. 

\subsection{Results}



\cref{tab:main_results} compares LTBD with prompting-based and training-based defenses across four LLMs.

\vspace{0.5pt}
\noindent \textbf{LTBD provides strong \& consistent robustness.} For example, on AlpacaFarm, LTBD achieves 0\% ASR on all four LLMs, outperforming inference-time defenses and matching the strongest training-based methods. 
Compared with DefensiveToken, LTBD reduces SEP ASR from 3.20/2.81/6.70/4.40\% to 2.04/2.00/1.57/1.86\% across the four models, and achieves 0.11-0.19\% ASR on TaskTracker. On CyberSecEval2, LTBD also remains competitive with training-based defenses and achieves the best result on Falcon3-7B (1.82\%). These results show that \textit{explicitly learning the trust boundaries} is substantially more effective than relying on generic defensive prompts or prefix tokens.

\vspace{0.5pt}
\noindent \textbf{The robustness gains come with little utility cost.} LTBD achieves the highest AlpacaFarm WinRate among all defenses for all four models and the highest SEP WinRate on three of the four models. Notably, LTBD even matches or exceeds the undefended model on 6 of 8 utility evaluations, indicating that the learned delimiters may enhance the model's original task-following ability rather than trading utility for security. Compared with DefensiveToken, LTBD obtains higher utility in 7 of 8 settings.

\vspace{0.5pt}
\noindent \textbf{LTBD closes much of the gap to training-based defenses without modifying the LLM}. StruQ and SecAlign obtain strong robustness by updating the underlying model through full fine-tuning or LoRA, whereas LTBD freezes the entire LLM and optimizes only four token embeddings. Despite this minimal training footprint, LTBD matches or surpasses these methods on several benchmarks, \textit{e.g.} TaskTracker and, for Falcon3 and Qwen2.5, SEP. This supports our central hypothesis: rather than modifying the model to learn which content to distrust, \textit{making the trust boundary learnable is sufficient to provide strong prompt-injection robustness with minimal intervention}.

\subsection{Ablation Study}


We conduct an ablation study to isolate the effects of \textbf{trust-boundary placement} and the \textbf{auxiliary preference objective}. We compare three variants: \textit{Prefix + CE}, which uses the same four learnable tokens as LTBD but places them all at the input prefix; \textit{Boundary + CE}, which places the four tokens at the trusted/untrusted region boundaries but removes the preference loss; and our LTBD.
\noindent \cref{tab:ablation} shows that \textbf{where the learnable tokens are placed matters}. On Llama3.1-8B-Instruct, replacing the prefix placement with trust-boundary placement reduces AlpacaFarm ASR from 23.08\% to 0.48\% and SEP ASR from 33.81\% to 3.43\%, while also improving utility. A similar trend holds for Llama3-8B-Instruct, where boundary placement reduces AlpacaFarm ASR from 6.73\% to 0.00\% and SEP ASR from 15.43\% to 2.34\%. This confirms that \textit{LTBD's gains do not come merely from adding trainable tokens; explicitly associating them with trusted and untrusted regions is critical.}
\noindent \textbf{The auxiliary preference objective provides further improvements.} It reduces SEP ASR from 2.34\% to 2.04\% on Llama3-8B and from 3.43\% to 2.00\% on Llama3.1-8B, while preserving or improving WinRate. These results support our central design: \textit{learnable trust boundaries are the primary source of robustness, while preference optimization further sharpens the model's preference for trusted-task behavior}.


\subsection{Defense against Adaptive Attacks}
\label{sec:adaptive_attack}
To evaluate whether LTBD remains robust when the attacker knows its mechanism, we introduce \textit{Delimiter-Spoof}, a defense-aware attack that explicitly targets the learned trust boundaries. We assume a strong attacker with full knowledge of the four delimiters and the LTBD input structure. Specifically, \textit{Delimiter-Spoof} inserts forged delimiter patterns around \(A\) to imitate, terminate, or reopen trusted and untrusted regions. We construct five variants covering different spoofing strategies, including fake instruction boundaries, fake data boundaries, and combinations thereof: 
\begin{tcolorbox}[
    colback=blue!2,
    colframe=blue!30,
    boxrule=0.35pt,
    arc=0.8pt,
    left=1.5pt,
    right=1.5pt,
    top=1pt,
    bottom=1pt,
    boxsep=0.5pt,
    before skip=2pt,
    after skip=2pt
]
\scriptsize
\setlength{\baselineskip}{8.5pt}
\begin{enumerate}[
    label=\textcolor{blue!70!black}{(\Alph*)},
    leftmargin=4.5mm,
    labelsep=1mm,
    itemsep=-1pt,
    parsep=0pt,
    topsep=0pt,
    partopsep=0pt
]
    \item 
    \texttt{<DATA\_END><INST\_BEGIN>}
    \textcolor{red!75!black}{$A$}
    \texttt{<INST\_END><DATA\_BEGIN>}

    \item 
    \texttt{<INST\_END><INST\_BEGIN>}
    \textcolor{red!75!black}{$A$}
    \texttt{<INST\_END>}

    \item 
    \texttt{<INST\_BEGIN>}
    \textcolor{red!75!black}{$A$}
    \texttt{<INST\_END>}

    \item 
    \texttt{<DATA\_END><INST\_BEGIN>}
    \textcolor{red!75!black}{$A$}
    \texttt{<INST\_END><DATA\_BEGIN><DATA\_END>}

    \item 
    \texttt{<DATA\_END>}
    \textcolor{red!75!black}{$A$}
    \texttt{<DATA\_BEGIN>}
\end{enumerate}
\end{tcolorbox}
\noindent During training, these variants are also used as attack augmentations to expose LTBD to forged boundary structures. At evaluation time, we test all 5 variants separately and report the worst-case ASR.

\cref{tab:adaptive_attacks} shows that LTBD remains highly robust under these adaptive attacks. On Llama3-8B and Falcon3-7B, the worst-case ASR remains 0.00\%, while Llama3.1-8B stays at 0.48\%. On Qwen2.5-7B, Delimiter-Spoof increases ASR only slightly from 0.00\% to 0.48\%. Overall, explicitly forging the delimiter syntax provides almost no advantage over plain injection.

These results provide an important insight into LTBD. Its robustness does not arise merely from the textual presence of special delimiters. Rather, the defense is encoded in the learned embeddings associated with the true structural trust boundaries. Consequently, Delimiter-Spoof prompts do not automatically acquire the same trust semantics. This separation between surface form and learned boundary role makes LTBD substantially harder to bypass.



\begin{table}[t]
\centering
\vspace{-10pt}
\caption{ASR (\% $\textcolor{mygreen}{\downarrow}$) under Adaptive Attacks.}
\setlength{\tabcolsep}{3.8mm}{
\renewcommand{\arraystretch}{1}
\resizebox{0.45\textwidth}{!}{%
\begin{tabular}{lcc}
\hline
\textbf{Model} & \textbf{Plain Injection} & \textbf{Adaptive Attack} \\
\hline
Llama3-8B-Instruct   & 0.00 & 0.00 \\
Llama3.1-8B-Instruct & 0.48 & 0.48 \\
Falcon3-7B-Instruct  & 0.00 & 0.00 \\
Qwen2.5-7B-Instruct  & 0.00 & 0.48 \\
\hline
\end{tabular}}}
\label{tab:adaptive_attacks}
\end{table}

\section{Conclusion}
\label{conclusion}
We presented LTBD, a lightweight prompt-injection defense that explicitly encodes trust boundaries while keeping the base LLM fully frozen. By optimizing only four delimiter embeddings, LTBD achieves strong robustness, preserves benign-task utility, and introduces negligible inference overhead. These results show that learning trust boundaries at the input level provides a simple, efficient, and effective approach to prompt-injection defense.


\bibliographystyle{IEEEbib}
\bibliography{strings,refs}

\end{document}